\documentclass[a4paper,12pt,reqno]{amsart}

\usepackage[centertags]{amsmath}
\usepackage{amsfonts}
\usepackage{amssymb}
\usepackage{amsthm}
\usepackage{newlfont}
\usepackage[centertags]{amsmath}
\usepackage{amsfonts}
\usepackage{amssymb}
\usepackage{amsthm}
\usepackage{newlfont}
\usepackage{color}

\theoremstyle{plain}

\theoremstyle{definition}

\theoremstyle{remark}

\numberwithin{equation}{section}

 \let\be=\beta  
   
 \let\la=\lambda \let\om=\omega

 \let\Ga=\Gamma

\newcommand{\caL}{{\mathcal L}}

\newcommand{\bbP}{{\mathbb P}}

\newcommand{\opunit}{\text{1}\kern-0.22em\text{l}}

\DeclareMathAlphabet{\mathpzc}{OT1}{pzc}{m}{it}

\newcommand{\beq}{ \begin{equation} }
\newcommand{\eeq}{ \end{equation} }

\begin{document}

\begin{center}
\noindent{\large \bf Symmetries of the ratchet current} \\

\vspace{15pt}

{\bf Wojciech De Roeck}\footnote{email: {\tt
wojciech.deroeck@fys.kuleuven.be}} and {\bf Christian Maes}\\
Instituut voor Theoretische Fysica, K.U.Leuven
\end{center}

\vspace{20pt} \footnotesize \noindent {\bf Abstract: } Recent
advances in nonequilibrium statistical mechanics shed new light on
the ratchet effect. The ratchet motion can thus be understood in
terms of symmetry (breaking) considerations. We introduce an
additional symmetry operation besides time-reversal, that switches
between two modes of operation. That mode-reversal combined with
time-reversal decomposes the nonequilibrium action so as to
clarify under what circumstances the ratchet current is a second
order effect around equilibrium, what is the direction of the
ratchet current and what are possibly the symmetries in its
fluctuations.

\normalsize

\vspace{20pt}

\section{Introduction}
Irreversible thermodynamics describes the appearance of currents
in macroscopic systems from specific nonequilibrium conditions.
The notion of entropy production is central and makes the product
of forces and fluxes. The forces are gradients of thermodynamic
potentials, directly connected to differences in concentration of
particles or to variations in temperature etc. The fluxes relate
to the transport of certain quantities. Basic information about
the direction of these currents follows from the second law of
thermodynamics (positivity of entropy production) and their
response and symmetry properties are contained in the Green-Kubo
and Onsager relations. Even though there is not yet a systematic
nonequilibrium theory beyond first order around equilibrium, for
many practical purposes that is not really problematic.\\

 The
situation is quite different and in fact, worse, for transport
phenomena that arise as rectifications of fluctuations such as in
Brownian motors \cite{reimannreview, astumian}. We will speak here
more generally about the ratchet effect. The very notion of
``ratchet effect" has not been uniquely defined in the literature,
perhaps witnessing the absence of a unifying understanding.
Yet, a few ideas are in common. It is e.g.\ emphasized that
ratchets are mesoscopic systems that provide transport in
spatially periodic media away from equilibrium, that ratchets are
driven by fluctuations and that the direction of transport cannot
be inferred from thermodynamics \cite{reimannhanggi}.

In the present paper we start from the idea that symmetry breaking
is central to the concept of ratchets. One is reminded of Curie's
principle that ``phenomena that are not ruled out by symmetries
will generically happen''. By symmetry, a sphere immersed in a
heat bath does not move. When one makes the object asymmetric, the
broken spatial symmetry does no longer inhibit directed motion.
However, if the heat bath is in equilibrium, the system still has
unbroken time-reversal symmetry (detailed balance) which prevents
motion. When finally also that time-symmetry is lifted, for
example by acting with a mixture of different baths at different
temperature, then the object will move. At least in principle,
since on macroscopic scales the effect will in general be blurred
by high inertia; the energy scales associated to the locomotion of
the object have to be
comparable with the thermal fluctuations induced by the
surrounding.

In what follows we contribute a general framework for ratchet
effects, based on symmetries of the action in the path integral.
Our main results are then as follows;

 First, we clarify when and
why the ratchet effect is second order. In a sense to be explained
the ratchet current is then orthogonal to
 the entropy production. As we will
specify,  that harmonizes well with the understanding that ``the
direction of the ratchet current does not follow from the Second
Law".  Secondly, we make the connection with the recently studied
fluctuation theorem. The ratchet work is in general the sum of
three physical quantities that each satisfy a fluctuation
symmetry.  Sometimes, but not always, the ratchet current itself
also satisfies a symmetry in its fluctuations. Finally, we discuss
how to infer the direction of the ratchet current.
 Of course, for specific models sharper bounds are
possibly available and the notions of ratchet work and of
efficiency can sometimes be discussed in much greater detail, see
e.g. \cite{parrondo2,vandenbroek}; in \cite{bier} one considers
explicitly second order currents and fluctuations of the ratchet
current have been studied in \cite{hanggi2}.  We emphasize however
that our work concerns general methods and tools in describing the
ratchet effect.  From a more fundamental perspective, it
illustrates and  exploits the role of the time-symmetric term in
the action governing the space-time histories of a system.  Our
analysis therefore takes part in the construction of
nonequilibrium statistical mechanics beyond the linear regime.

\section{Ratchet essentials}
We start by explaining our particular point of view on ratchet
systems.
\subsection{Fluctuations} Ratchet devices are best described on a microscopic or
mesoscopic scale where in the usual set-up one considers
stochastic processes as specified from some master or kinetic
equation.  We do not need a specific model equation (but we will
be giving examples below) and we assume that for the appropriate
scale of description the distribution of histories is given after
some transient time as weighted via some generalized
Onsager-Machlup Lagrangian $\caL_{\la}$ \beq\label{pathm} \mbox{
Prob}[\omega] \propto e^{- \caL_{\la}(\om) }\,\bbP^0(\om) =:
\bbP^\la(\om) \eeq We explain the notation.  The $\om=(\omega_t)$
are paths or histories of the system over a certain time-interval
$[0,T]$, where at each time $t$, $\om_t$ describes the state of
the device. The weights of $\om$ are given in terms of the
functional $\caL_{\la}$, called the action or the Lagrangian,
extensive in the duration $T$ (not explicitly indicated for
simplicity of notation). All quantities derived from the
Lagrangian $\caL_{\la}$ are only defined modulo a temporal
boundary term, i.e., a difference of the form $U(\omega_T) -
U(\omega_0)$, and below, we often write equalities between
functions of paths $\om$, which would be incorrect if we did not
allow for such a boundary correction.

 In the case of small
macroscopic fluctuations, the $\caL_{\la}$ is known as the
Onsager-Machlup Lagrangian.  More generally, it is simply obtained
by taking a path integral representation, i.e., taking the
logarithm of the path-probabilities as from discrete time
approximations or from so called multi-gate probabilities or from
a Girsanov formula for Markov processes, see e.g. \cite{oksendal}.

 The Lagrangian $\caL_{\la}$ depends on a parameter
$\la$ which represents a particular driving that will generate the
ratchet current. For $\la=0$, the process $\bbP^0$ is a reference
process; we assume that all the nonequilibrium driving resides in
$\caL_{\la}$ so that $\bbP^0$ is in fact a corresponding
equilibrium process. Nonequilibrium expectations are computed with
the nonequilibrium path-space distribution \eqref{pathm}
\[
\langle f \rangle_\la = \int d\bbP^0(\om)
\,f(\om)\,e^{-\caL_\la(\om)}
\]
for the normalized expectation of a function $f(\omega)$ in
histories $\om$.\\

\subsection{Modes of operation}
A ratchet device can be considered as a motor that has available
various different pathways or channels to complete its working
cycle.  In  general, the state of the ratchet is represented by
two coordinates; $x$, a 1-dimensional cyclic coordinate which
gives the position of the motor and $k$, mostly discrete and which
specifies additional information. If $k$ can take only one value,
then the motor has basically only one pathway; if $k$ takes more
values (we restrict ourselves to two values), then the motor can
switch the $k$ coordinate during its cycle, and hence there will
be different types of channels or pathways. No explicit
thermodynmic force needs to be specified.  The coordinate $k$ can
be spatial (e.g.\ like in Feynman's ratchet and pawl), it can
determine the type of environment (when the motor interacts with a
gas consisting of multiple species which are not in equilibrium
with each other), it can specify the potential (like in a flashing
ratchet) or the value of some time-dependent external field.  In
some cases, the different modes of operation could represent
different energy levels of the system and the switching then
results from contact with a heat bath, cfr.\ thermoelectric
effects as in \cite{maesvanwieren}.  In summary, the paths $\om$
we have in mind when writing \eqref{pathm} also include the
information what temperature, or what potential etc.\ is used
($k$-coordinate) at what time, and not only the position of the
motor itself ($x$-coordinate).

Since $k$ takes two values, these channels can be divided in a set
of pairs, and we can usefully define a transformation between the
two members of the pair. More generally and for each path $\omega$
we can associate to it a transformed path $\Gamma \omega$,
obtained by switching $k$'s in each step of the path and thus
switching the modes of operation of the motor.
  We emphasize that  the
symmetry $\Ga$, called mode-reversal, acts directly on path space
as we include in the history the setting of the driving or of the
environment. E.g. $\Ga$ allows to exchange two different
potentials or temperatures etc. One can have in mind that $\Ga$ is
(effectively) a sign-reversal of the thermodynamic forces, e.g
$\lambda \rightarrow -\lambda$. In the case of devices with
external periodical forcing, $\Ga$ corresponds to shifting each
path by one
half of the period of the external force.\\

Besides mode-reversal and as essential in all nonequilibrium
systems one can also apply time-reversal.  One then compares the
weight of a trajectory $\omega$ with that of its time-reversal
$\theta \omega$: $(\theta \omega)_t = \omega_{T-t}$. We restrict
us to variables like particle positions, and we do not consider
here variables that are odd under kinematical time-reversal (like
velocities).   The difference between the probabilities for
$\omega$ and $\theta \omega$ measures the irreversibility, as has
been expressed in a number of fluctuation relations over the last
years, see
\cite{maesoriginanduse} for a review.\\

It is the breaking of the $\Ga-$symmetry, combined with breaking
of detailed balance, that generates the nonequilibrium ratchet
effect. It generates a nonzero ratchet current $J_r$ measuring the
cycling speed, at least when there are no further symmetries that
would forbid $J_r\neq 0$.  We now consider the symmetry properties
of the path-dependent ratchet current $J_r$. In contrast with many
situations close to equilibrium,
  we need to introduce yet
other considerations than strictly related to entropy production
or time-reversal (breaking).   Now comes the relevance of the
symmetry operation $\Ga$. We say that $J_r$ is a {\it ratchet
current} (associated to the operation $\Ga$) if it satisfies both
\begin{eqnarray}\label{cond: symmetries}
J_r(\Ga \om) &=& J_r( \om) \\
J_r(\theta \om) &=& -J_r( \om) \nonumber
\end{eqnarray}
The first symmetry of $J_r$ under $\Gamma$ means that the ratchet
current simply counts the number of completed cycles (in the
$x$-coordinate) no matter along what channel (choices of
$k$-coordinate) it was taken; as a current  counting the steps of
the ratchet in $\omega$ we  naturally ask that $J_r(\omega)$ is
antisymmetric under time-reversal $\theta$.

\section{First order vs. second order}\label{sec: first vs second}

 We
require that the equilibrium situation is $\theta-$symmetric
 \beq
\label{cond: eq is reversible} \bbP^0(\theta\om)= \bbP^0(\om)
 \eeq
which implies that in equilibrium $\langle J \rangle_0=0$ for all
time-antisymmetric observables $J$. The nonequilibrium driving
breaks the time-symmetry and we let $S_\la =S$ be the
$\theta-$antisymmetric part of the Lagrangian, i.e.,
 \beq \label{defs}
 S= \caL_\la(\theta \om)-  \caL_\la (\om)
 \eeq
  It turns out that the variable $S$ can be
identified with the path-dependent entropy production appropriate
to the scale of description, \cite{maesnetocny,maesoriginanduse},
always up to a total time-difference. Obviously, $S(\theta \om)=
-S( \om) $.

\subsection{Orthogonality}
For ratchets it is very useful to employ also the mode-reversal
$\Ga$, and to put $\om$ in the balance {\it versus} $\Ga\om$.
 To start we also ask here that \beq \label{ents} S(\Ga\om)= -S(
\om) \eeq
 which is straightforward in most concrete
models (think e.g.\ of heat conduction where one exchanges the
temperatures of baths for a fixed history $\omega$).
 Remark that the entropy production $S$ and the
ratchet current $J_r$ then behave differently under the symmetry
$\Ga$, but identically under the symmetry $\theta$.\\

 Clearly, from the properties $S\Gamma
= -S,
 J_r\Ga = J_r$ follows that the mutual covariance between $S$ and
 $J_r$ equals zero
 \beq \label{eq: ortho
with a priori}
 \int Q(d \om)\, J_r (\om) S(\om ) = 0,\quad
 \int Q(d \om)\, S(\om ) = 0
 \eeq
for no matter what $\Ga-$invariant distribution $Q$. The identity
\eqref{eq: ortho with a priori} expresses an orthogonality or
independence between the variable entropy production and the
ratchet current. It announces that the ratchet effect plays beyond
irreversible thermodynamics and there arises for example the
problem of determining the direction of the ratchet current.\\

One can indeed learn something about the ratchet effect by the
usual perturbation theory around equilibrium. One then expands the
nonequilibrium state $e^{-\caL_\la }\bbP^0$ around equilibrium
$\bbP^0$ to obtain, via \eqref{cond: eq is
reversible}-\eqref{defs},
 \beq\label{resp}
 \langle J_r\rangle_\lambda =
 \frac 1{2} \langle J_r S_\la \rangle_0 +O(\la^2)
\eeq The consequence of \eqref{eq: ortho with a priori} now
appears. In many cases, including almost all flashing ratchets,
the equilibrium process is invariant under $\Ga$.  Then we can
take
 $Q (\om)= \bbP^0(\om)$ in \eqref{eq:
ortho with a priori} and $\langle J_r S_\la \rangle_0=0$. As a
result, from \eqref{resp} we see that the ratchet current vanishes
in first order in $\la$. The reason is the invariance of the
equilibrium process under $\Ga$ combined with the antisymmetry of
the entropy production $S$ under $\Ga$.  That appears to be the
general mechanism when obtaining ratchet effects only in second
order around equilibrium.
At the same time, we see that first order ratchets appear  when
the equilibrium state $\bbP^0$ is not $\Ga-$invariant; see
\cite{jarzynskimazonka} for a simple example.

\subsection{Ratchets with load} 
When one attaches a load to extract work from the ratchet effect,
the above description must be modified.  Applying a load is
effectively coupling the ratchet current to the entropy
production. It is now no longer true that the entropy production
$S$ is antisymmetric under $\Ga$ and the relation \eqref{ents} no
longer holds.  To further resolve the (anti-)symmetries, we
decompose $S_\lambda$ into
\[
S_\lambda = S^+_\lambda + S^-_\lambda
\]
 where $S_\la^+= S_\la^+\Ga$ ($S_\la^-= -S_\la^-\Ga$) is
(anti-)symmetric under $\Gamma$. As an example, we can already
think of a heat engine working between inverse temperatures
$\beta_1$ and $\beta_2$.  The variable entropy current is $S =
\beta_1 J_1 + \beta_2 J_2$ where $J_i$ the heat current into
reservoir $i$, while the delivered work equals $-W = J_1+J_2$
(energy conservation). Then, \begin{equation}\label{sent} S =
\frac{1}{2}(\beta_1 - \beta_2)\, (J_1 -J_2)+ \frac{1}{2}(\beta_1 +
\beta_2)\, W \end{equation}

 We think of the exchange of heat
baths as a mode-reversal and  we can take $ \lambda \sim \beta_1 -
\beta_2$. The first term in \eqref{sent} is antisymmetric  under
the exchange $\beta_1 \leftrightarrow \beta_2$ and the second term
(containing the work $W$) is symmetric under $\Ga$. Quite
generally, the term $S^+$ turns out to be proportional to the work
done on the ratchet, as function on path-space.
Assuming that ratchet work is proportional to the number of
completed cycles (as can be checked quite often) we  write the
work as $S^+= -f J_r$ for a constant load $f$. As a consequence,
the linear term in \eqref{resp} gets rewritten as
\[
 \langle J_r \rangle_{\lambda,f} =
 \frac 1{2}\langle J_r S^-_\lambda \rangle_0
  - \frac 1{2} f\,\langle J_r J_r \rangle_0 +
 O(\lambda^2,f^2)
 \]
Again, the first term on the right (coupling heat dissipation with
the ratchet current) vanishes if the equilibrium state $\bbP^0$ is
$\Ga$-invariant and the response of the ratchet current to the
load is in first
order determined by a current--current autocorrelation (the second term on the right). 

\section{Examples}

Ratchets allow motion without the application of net thermodynamic
forces. The difference between a ratchet and a {\it perpetuum
mobile of the second kind} arises from the nonequilibrium
condition. Depending on the specific nature of the nonequilibrium
one distinguishes different kinds of ratchets.  As a result the
above notions are realized in a somewhat different way for
flashing ratchets, rocked ratchets, Feynman ratchets,
B\"uttiker-Landauer ratchets etc.  To fix the ideas and to
illustrate the basic concepts, we consider here two classes of
ratchet systems.

\subsection{Two-temperature ratchet}\label{sec: twotemp}

A particle travels  on a periodic landscape, modeled by a double
ring whose sites are indexed by $(x,k)$ with $x=0,\ldots,L$  and
$k=1,2$. Site $0$ is identified with $L$. An asymmetric potential
function $V(x)$ is given. In each step the particle can either
jump from $(x,k)$ to $(x\pm 1,k)$, or it can change its
$k$-coordinate while keeping $x$ unchanged. One could have in mind
that the particle moves on the interface between two gas
reservoirs; whenever $k=1$, it interacts with reservoir $1$ and
analogously for $k=2$. The reservoirs have respective inverse
temperatures $\be_{1,2}$. The dynamics is given by a Markov jump
process with jump rates

 \beq \label{jr}
 c((x,k),(y,k)) =  g_{k}(x,y)
\,e^{-\be_{k} (V(y)-V(x))/2}
 \eeq
for jumps from $x$ to a nearest neighbor $y=x\pm 1$ on the ring,
and
 \beq \label{jr2}
 c((x,k),(x,k')) =  c((x,k'),(x,k)) =   h(x)
 \eeq
for a change of $k \rightarrow k'$. In going from $(x,k)$ to
$(y,k)$, the particle absorbs energy $V(y)-V(x)$ from reservoir
$k$.  We demand that $g_{k}(x,y)=g_{k}(y,x)$ and the symmetry
\eqref{jr2} to assure that the only source of entropy creation in
the jump is by the transfer of heat $V(y)-V(x)$ (see also the
first paragraph of Section \ref{sec: first vs second}). The
functions $g_{k}(x,y)$ can for example include details about the
chemical potential of the reservoir, or more generally, about the
contact between the reservoir and the particle. Remark that an
eventual chemical potential does not cause any entropy production
since no gas particles are being transported between the two
reservoirs.

 The driving $\lambda$ can then be identified with
the difference between the two reservoirs, say in terms of
$\beta_1-\beta_2$ and $g_1(x,y)-g_2(x,y)$.  We make hence the
assumption that $g_1(x,y)=g_2(x,y)$ when $\beta_1=\beta_2$,
corresponding to equilibrium. The paths $\om$ correspond to
sequences of positions $x_j,k_j$ and of jump times $t_j$:
\beq\label{flapa}
 \om = (x_1,k_1,t_1; x_2,k_2,t_2; \ldots; x_n,k_n)
\eeq Time-reversal $\theta$ (for some large $T$) transforms the
path $\omega$ into $\theta \om= (x_n,k_n,T-t_{n-1},k_{n-1};\ldots;
x_2,k_2,T-t_1; x_1,k_1) $.  The mode-reversal $\Gamma$ exchanges
the reservoirs and it works on the $k_j$'s exchanging $k=1,2$ The
two reservoirs are identical in the equilibrium process
($ \lambda=0 \Rightarrow \beta_1=\beta_2, g_1(x,y)=g_2(x,y) $).\\

The antisymmetric term \eqref{defs} under time-reversal in the
Lagrangian can be obtained from computing \[S(\om) =\log
\frac{\bbP (\om) }{\bbP (\theta \om)} = \log \frac{c((x_1,k_1),
(x_2,k_2) )\ldots
c((x_{n-1},k_{n-1}),(x_{n},k_{n}))}{c((x_{n},k_{n}),(x_{n-1},k_{n-1}))\ldots
c( (x_2,k_2),(x_1,k_1) )} \] or
 \beq \label{ss}
  S(\omega) =
\sum_{j=1}^{n-1} \be_{k_j} \big(V(x_{j+1})-V(x_{j})\big) \eeq
which is the sum of changes in the entropy of the gases (Note that
the jumps where the $k$-coordinate changes, do not enter
$S(\om)$). The particle itself is thought of as microscopic and
not contributing to the entropy, so that \eqref{ss} is the
path-dependent entropy production.

 Clearly, $S$ is antisymmetric under time-reversal.
There is another way of writing \eqref{ss} to make clear that $S$
is also antisymmetric under $\Gamma$: \begin{align}
 S(\omega) = -\beta_1 \sum_{j:k_j=1} \big(V(x_{j+1})-V(x_{j})\big)
- \beta_2 \sum_{j:k_j=2}\big(V(x_{j+1})-V(x_{j})\big)&&\nonumber\\
= -(\beta_1-\beta_2) \sum_{j:k_j=1} \big(V(x_{j+1})-V(x_{j})\big)
- \beta_2 \big(V(x_{n})-V(x_{1})\big)&&\nonumber\\
= - (\beta_1-\beta_2) \sum_{j:k_j=1}
\big(V(x_{j+1})-V(x_{j})\big)&&
\end{align}
The last equality illustrates our convention that all
path-dependent
quantities are written modulo a total time-difference.\\

Clearly, the ratchet current $J_r(\omega)$ is a function of
$\tilde \omega = (x_1,t_1;\ldots;x_n)$ only and it does not depend
on the $k_j$'s.   Its mean $\langle J_r \rangle$ is generically
nonzero when $V$ is asymmetric (and no other accidental symmetries
are present). The ratchet is second order (this is due to our
assumption that $g_1(x,y)= g_2(x,y)$ when $\be_1=\be_2$); the
entropy
production \eqref{ss} is not of the form ${\cal F}\,J_r$.\\


\subsection{Flashing ratchet}\label{sec: flash}

In the previous example, it was the environment (and specifically
the temperature) that was effectively changing between two
possible values. We can also take the time-dependence in the shape
of the potential. As another difference we consider now a Langevin
set-up.  Again it concerns a second order ratchet.

Consider  a particle  in a spatially periodic landscape with the
potential flashing between two potential functions $V_{+1}$ and
$V_{-1}$, both periodic functions
$V_{\pm 1}(x)=V_{\pm 1}(x+L)$. Again, one has to eliminate
additional symmetries, like mirror symmetry of the potentials or
supersymmetry  \cite{reimannreview}, to get a nonzero ratchet
current.

The nonequilibrium parameter $\la$ measures the difference between
the two potentials parameterized as $V_{\pm 1}=V \pm \la W$.
 The particle is in contact
  with a heat bath at inverse temperature
   $\beta$. We model its motion by the overdamped
    Langevin equation
\beq \label{eq: overdamped}  \dot x_t = -V'_{k(t)}(x_t) \, + \xi_t
\eeq where
$\xi_t$ is a fluctuating Gaussian force with white noise
statistics: $ \langle \xi_t \rangle =0 $ and $\langle \xi_s \xi_t
\rangle = 2 \beta^{-1}\delta(t-s)$.  The time-dependence $k_t=\pm
1$ is arbitrary.
   The
reference process has $\la=0$, meaning that the potential is fixed
equal to $V$. Under It\^o-convention, one shows
\beq \caL_\la=\frac{\beta}{2}\big[\lambda \int dx_t k_t\,W'(x_t) +
\lambda \int dt\, k_t\,V'(x_t)\,W'(x_t) + \frac{\lambda^2}{2} \int
dt\,W'^2(x_t)\big]
 \eeq
 The paths are given as $\omega_t= (x_t,k_t)$
with time-reversal implemented by (for some large $T$) $\theta
(x_t,k_t)= (x_{T-t},k_{T-t})$
 and \beq S = \caL_\la \theta-\caL_\la =
-\beta\,\lambda \int dt \,\dot{k_t}\, W(x_t)\eeq which is $\beta$
times the dissipated power through the external forcing.
 The mode-reversal $\Ga$  switches potentials: $
\Ga (x_t,k_t)=(x_t,-k_t) $ and one observes that $S\Gamma=-S$.

\section{Ratchet fluctuations}\label{ratf}
 The two symmetry operations $\Ga$ and
$\theta$ suggest a natural decomposition of the Lagrangian
$\caL_\la$. From now on we assume that $\theta\Ga = \Ga \theta$
(commutativity\footnote{That is generally true if the ratchet
coordinate can be separated as $(x,k)$, $\Ga$ acts by changing $k$
and $\theta$ does not mix $x$ and $k$.  Both examples in Section 4
have that property.} ). We write \beq \label{def: R} R =R_\la =
(\caL_\la\theta\Ga + \caL_\la\Ga - \caL_\la\theta
 - \caL_\la)/2
 \eeq
 for the part that is antisymmetric under $\Ga$ and is symmetric
 under $\theta$.
 The
Lagrangian has the form \beq\label{formL}
 \caL_\la = \caL^{+}_\la -\frac 1{2}[R_\la + S_\la]
\eeq where $\caL^{+}_\la$ is $(\theta,\Ga)-$invariant.

One can now verify that ratchet models typically satisfy various
fluctuation theorems. In brief,
 when $\bbP^0$ is
$(\theta,\Gamma)-$invariant, then for all the three choices
$V=S,R+S^+,R+S^-$,
\begin{equation}\label{ft}
\frac{\bbP^\la(V=v)}{\bbP^\la(V=-v)} = e^{v}
\end{equation}
For $V=S$, \eqref{ft} is similar to the Gallavotti-Cohen
fluctuation symmetry for the fluctuations of the entropy
production \cite{gallavotticohen1,gallavotticohen2}; for
$V=R+S^+$, \eqref{ft} has been derived in
\cite{maesvanwierensymmetric}; finally, \eqref{ft} also holds for
 $V=R + S^-$.  The reason why in all these
 cases, one finds that fluctuation relation
is that $S,R+S^-$ and $R + S^+$ are the antisymmetric parts in the
Lagrangian $\caL_\lambda$ under respectively the symmetries
$\theta,\Ga$ and $\theta \Ga$. The relation \eqref{ft} can in each
of the three cases be directly verified from computing the ratio $
\bbP^\la(\om)/\bbP^\la(Y\om)$ for transformations $Y=\theta, \Ga,
\theta \Ga$ in \eqref{pathm}, and from combining that with the
decomposition \eqref{formL}. In order to control temporal boundary
terms, it is assumed that the system itself has a bounded state
space; otherwise, some extended fluctuation symmetry can be
expected, see \cite{baiesijacobsmaes,vanzoncohen}.

Observe also  that the ratchet work $S^+ = \big[S +( R+S^+) - (R +
S^-)\big]/2$ is a sum of three observables, each of which
satisfies a fluctuation theorem \eqref{ft}.

 A natural question is whether
the ratchet current $J_r$ itself satisfies a fluctuation symmetry.
In general, the answer seems to be negative, but nevertheless it
is possible to construct classes of ratchets where that symmetry
is verified, as is also remarked for some specific models in
\cite{andrieuxgaspard,sakaguchi}, and as now will be shown.

We come back to the 2-temperature ratchet of Section \ref{sec:
twotemp}. We consider the limiting case of a very rapid changing
of the reservoir ($k$-coordinate), hence the limit $h(x) \uparrow
+\infty$. Another possible realization is obtained by thinking of
the particle as a rigid body extended and connected at its ends to
two different reservoirs.  Then, we have a simple model of the
Feynman-Smoluchowski ratchet much in the spirit of
\cite{vandenbroekkawai} but in the overdamped limit.  With respect
to \eqref{flapa}, we make a more coarse grained description and we
only look at the particle jumps (forgetting about what reservoir
caused it), i.e., the jump rates are now between $x$ and $y$ and
they are given by the sum $c(x,y)= c((x,1),(y,1))+ c((x,2),(y,2))
$.  In other words, we collect several of the original paths $\om$
of \eqref{flapa} into one and the same new path $\tilde{\om} =
(x_1,t_1;x_2,t_2;\ldots; x_n)$. Obviously now the $\Ga$-symmetry
has left the stage and there is effectively only one possible
channel (though of course, if one wants to keep track of the
physical entropy production, one still has to distinguish which
reservoir ``caused'' what transition). The corresponding pathspace
distribution is
 \beq\label{aj}
 \tilde{\bbP}(\tilde\om) = \sum_{\omega\rightarrow
 \tilde\omega}\bbP^\la(\om),\quad
  \tilde\bbP(\tilde\om) \propto e^{-
\tilde{\caL}(\tilde\om)}  \eeq with a new Lagrangian
$\tilde{\caL}$. The key observation is that pathwise, its
antisymmetric component $\tilde{\caL}(\theta \tilde\om) -
\tilde{\caL}(\tilde\om) $ is proportional  to  the ratchet current
\beq a J_r  =   \tilde{\caL}\theta  - \tilde{\caL} \eeq with a
constant $a$ that can be computed explicitly. By standard
arguments it follows that
 \beq\label{fsf}
\frac{\bbP^{\la} (J_r =  j )}{\bbP^{\la} (J_r = - j )} =
  e^{ a j} \eeq
  which is a fluctuation symmetry for the ratchet current.
In particular   $a\langle J_r\rangle_\la \geq 0$, which obviously
determines the sign of the ratchet current.

Note that in this limit, there is a new accidental symmetry
possible; if $g_1(x,y)=g_2(x,y)$ for some nonzero  $\la$, then one
easily checks that $c(x,y)/c(y,x)=e^{A(y)-A(x)}$ for some function
$A$. This ``effective" detailed balance condition immediately
implies $J_r=0$. The same kind of symmetry can be seen in the
ratchet \cite{vandenbroekkawai} if one models the contact with the
thermal baths by Langevin forces (instead of a Boltzmann equation,
as in done in \cite{vandenbroekkawai}).

\section{Direction of ratchet currents}\label{dirr}

 In first-order ratchets,
one can interpret \eqref{resp} as a principle for determining the
direction of the ratchet current close to equilibrium, providing a
simple mathematical explanation of the ideas in \cite{komatsu}.
Indeed, since ${\bbP}^0 [J_r
>0]={\bbP}^0 [J_r <0]$, we can evenly split
\begin{equation}
\langle J_r S_\la \rangle_0 = 1/2 \langle J_r S_\la | J_r >0
\rangle_0 +1/2\langle  J_r S_\la | J_r <0 \rangle_0
\end{equation}
Combine that with the fact that  $\langle S_\la \rangle_0=0$ to
conclude that if the entropy
production $S_\la$
 is overwhelmingly positive in one of the
two subensembles $J_r >0$ or $J_r <0$, then the ratchet current
has the sign as in that subensemble.

 For more general ratchets, one can use the consequences of
the fluctuation theorems \eqref{ft}. It implies that $S,R+S^+$ and
$R + S^-$ are all positive with a probability that exponentially
approaches $1$ as the duration $T \uparrow \infty$. In principle,
that determines
the direction of the ratchet current.\\
To be more specific we consider unloaded ratchets for which the
first order around equilibrium vanishes, see the discussion around
\eqref{resp}. Then, the first non-vanishing order is given by
 \beq\label{resp2}
 \langle J_r\rangle_\lambda =
 \frac 1{4} \langle J_r S_\la R_\la \rangle_0 +O(\la^3)
\eeq Hence, one has to study the sign of
 $S_\la R_\la$ in the two equilibrium subensembles
 $J_r >0$ and $J_r <0$.
Typical trajectories are characterized by having positive entropy
production $S_\la>0$.  Yet, that does not yet fix the direction of
the ratchet current in the case of second order. The
time-symmetric term $R_\la$ must however also be positive for
typical paths.  That selects within the class of paths where
$S_\la>0$ what the direction of the current will be.\\



\section{Conclusions}

  Fluctuations are driving the ratchet effect. It is
therefore important to investigate the structure of the action in
the path-integral governing the path-probabilities.  Another
symmetry transformation $\Ga$ (mode-reversal) appears that
together with time-reversal decomposes the nonequilibrium action.
The term in the Lagrangian action that is symmetric under
time-reversal but is antisymmetric under mode-reversal,
contributes significantly to determining the direction and the
nature of the fluctuations of the ratchet current.  That effect is
most outspoken for second order ratchets.\\

\noindent {\bf Acknowledgments:}
 We thank P.~Reimann and K.~Neto{\v{c}}n\'y
for valuable suggestions. WDR is a postdoctoral fellow of the FWO
(Flanders).  C.M. benefits from the Belgian Interuniversity
Attraction Poles Programme P6/02.


\bibliographystyle{plain}
\bibliography{ratchetPRE4}

\end{document}